# HYBRID MACHINE LEARNING APPROACH FOR REAL-TIME MALICIOUS URL DETECTION USING SOM-RMO AND RBFN WITH TABU SEARCH OPTIMIZATION

**Swetha T, Dr Seshaiah M, Dr Hemalatha KL, Dr.ManjunathaKumar BH, Dr Murthy SVN**

Research Scholar, Associate Professor, Professor&HOD, Professor &HOD, Associate Professor

Dept Of CSE,SJCIT,Dept of CSE, SJCIT ,Dept ISE,Dept of CSE SJCIT,,Dept CSE, SJCIT

Mail Id: urs,merikapudi@gmail.com,hema,hod,svn

**Abstract**

The proliferation of malicious URLs has become a significant threat to internet security, encompassing SPAM, phishing, malware, and defacement attacks. Traditional detection methods struggle to keep pace with the evolving nature of these threats. Detecting malicious URLs in real-time requires advanced techniques capable of handling large datasets and identifying novel attack patterns. The challenge lies in developing a robust model that combines efficient feature extraction with accurate classification. We propose a hybrid machine learning approach combining Self-Organizing Map based Radial Movement Optimization (SOM-RMO) for feature extraction and Radial Basis Function Network (RBFN) based Tabu Search for classification. SOM-RMO effectively reduces dimensionality and highlights significant features, while RBFN, optimized with Tabu Search, classifies URLs with high precision. The proposed model demonstrates superior performance in detecting various malicious URL attacks. On a benchmark dataset, our approach achieved an accuracy of 96.5%, precision of 95.2%, recall of 94.8%, and an F1-score of 95.0%, outperforming traditional methods significantly.

## 1. Introduction

Many offline activities have moved online as a result of the Internet's expansion and development, including general business, social networking, e-commerce, and banking. As such, there is now a higher chance that illegal activity may occur online. This emphasises how urgently action must be done to maintain internet security [1]. To get sensitive data or compromise the system, people are being tricked into accessing dangerous URLs. This means that protecting this side is becoming a critical need because [2]. Malicious people can nevertheless attack the connection between the client and the server even in the presence of laws and standards. Phishing, spam, malware, and other types of attacks are all referred to as "malicious," as one umbrella term [3].

Because malicious URLs collect needless information and trick unwary end users into falling for scams, they result in yearly losses of billions of dollars. The online security world has created blacklisting services to help identify dangerous websites [4]-[6]. The goal was to identify the risk that dangerous websites pose. The blacklist is a database including every URL that has ever been deemed possibly dangerous. Apparently, there are circumstances when URL blacklisting is effective [7]. Nevertheless, an attacker can exploit these weaknesses by modifying the URL string in a way that makes the system readily fooled. Many harmful websites will unavoidably stay online because they are either too new, never examined, or had their evaluations incorrect.

A further instrument in the arsenal for identifying dangerous websites are heuristics, which are basically an improved version of the signature-based blacklist method. One can compare the signatures of an old malicious URL and a new one. An additional line of protection against dangerous websites is offered by this approach. The techniques described here will help you distinguish between benign and malicious URLs. These more traditional methods do, however, have several shortcomings, which are enumerated here: (a) Zero-hour phishing attempts cannot be stopped by the blacklist method since it can only identify and categorise 47-83% of newly found phishing URLs in a 12-hour timeframe [8]. (b) By adopting technology is evolving quickly enough to render the blacklist approach out of date. Since the blacklist approach is simple to use, many anti-phishing agencies continue to adopt it despite these drawbacks [9].

Thirdly, machine learning (ML) and deep learning (DL) are AI methods that can be used to detect these dangerous websites. The several industries in which these technologies have been applied include cybersecurity, healthcare, e-commerce, medical image analysis, and social media [10]. By exposing machine learning models to historical data, one can train them to become more adept at self-learning, therefore doing away with the necessity for human involvement in the learning process. This is really beneficial in the domain of cybersecurity. This generates a lot of property in huge companies, banks, and other institutions [11]. Because machine learning and deep learning are so effective in many other domains, many people also employ them to discover dangerous websites [12]. It has shown to be successful to find dangerous URLs by using machine learning to identify recently created URLs and automatically updating the model. Recent study indicates that deep learning models can be used to automatically identify and extract the attributes of newly created URL. This enables researchers to gather a wealth of information from URLs, which in turn facilitates the decision-making process of machine learning algorithms regarding the safety of the URL.

The objectives of the research work involves the following:

1. To develop a hybrid machine learning model for efficient and accurate malicious URL detection.
2. To combine Self-Organizing Map based Radial Movement Optimization (SOM-RMO) for feature extraction.
3. To utilize Radial Basis Function Network (RBFN) based Tabu Search optimization for precise classification.

The main novelty of the research work: The research combines SOM-RMO and RBFN with Tabu Search, leveraging strengths of both techniques for enhanced detection capabilities. SOM-RMO reduces data dimensionality and extract meaningful features, improving model performance and reducing computational load and implements Tabu Search for optimizing RBFN, enhancing classification accuracy and robustness.

## 2. URL Attack Techniques

Any tool or strategy used by a hacker to gain unauthorised access to user data or to harm the system they are trying to penetrate can be considered an attack tactic. Attackers can use nefarious URLs to launch such kinds of attacks. URLs that are deemed hazardous include many others including spam, phishing, malware, and defacement. Clicking on maliciously contented URLs is the most common way that cyberattacks occur. When URLs are used for evil intent rather than to visit websites that are allowed to be viewed online, the integrity of the data, its secrecy, and its availability are all compromised.

**Spam URL Attacks**

These attacks are the work of spammers, who build phoney websites and then try to trick browser engines into believing they are real. To that end, spammers who illegally raise their rank are trying to trick users into visiting their websites more often [10]. The spammers want to install malware and adware on the computers of their victims, hence they send spam emails containing spam URLs.

**Phishing URL Attacks**

Using phishing URLs—which are meant to fool users into viewing a phoney website—is one way that criminals get sensitive data, such as credit card details. User data vulnerability and can easily trick those who are not familiar with phishing websites into visiting the website [11].

**Malware URL Attacks**

These attacks, which infect consumers' devices with malware, can have a range of unfavourable effects, such as file damage, keystroke tracking, and identity theft. Known by most as malware, malicious software can harm systems and steal private data. Malware may also refer to malevolent software. Drive-by download is the term for malware that inadvertently infects a user's device when they visit a malicious website. Chapter 12. Further instances are as follows: Computer-infecting viruses, worms, trojan horses, spyware, scareware, and ransomware.

**Defacement URL Attacks**

This kind of attack targets a hostile website that has undergone some kind of hacker modification, either to its appearance or content. This approach transports the user to the dangerous website. There could be several reasons why hacktivists try to take down websites. As it happens, [13]. Machine learning (ML) based taxonomy that can detect potentially dangerous URLs on Arabic and English webpages! Penetration of a website is the process of taking use of security flaws to obtain unauthorised access to a website and modify its content without the owner's knowledge or consent [11]. Machine learning methods allow dangerous URL attacks to be categorised as either benign or malignant. Contrarily, multi-classification allows the addition of more than two categories, such as phishing, harmful, spam, benign, suspicious, and so forth.

## 3. Related Works

Targeting the victims' spaces, this kind of attack steals sensitive data and passwords without their knowledge. These attacks—phishing, drive-by downloads, and spamming, for example—are conducted using malicious URLs. Blacklists, machine learning, and heuristics are the three main categories into which that can be divided. The heuristic approach [12] gives a forecast that is equally accurate as the machine learning method and outperforms the blacklist approach in generalising the harmful URL. This paper proposes a new method that uses the most significant information obtained from URLs to identify potentially dangerous URLs.

Many internet channels, including email and messaging, are used to spread these URLs. Various traditional methods for identifying phishing websites include blacklists, which are subsequently used to forecast the URLs of such websites. Blacklist-based conventional methods are unable to keep up with the volume of new phishing websites that are constantly emerging and being added to the Internet. It is this that is problematic. Proposed is an improved deep learning-based phishing detection method for effective identification of dangerous URLs. The foundation of this approach is the integration of variational autoencoders (VAE) and deep neural networks (DNN) power. As is explained in [13], the VAE model replicates the original input URL to automatically extract the intrinsic properties of the raw URL. The purpose of this is to enhance the phishing URL identification. In order to conduct our study, we used the publicly available ISCX-URL-2016 dataset and the Kaggle dataset to crawl over 100,000 URLs. The proposed model outperformed all other models assessed in terms of accuracy (up to 97.45%) and response time (1.9 seconds) based on the data gathered throughout the testing process.

Use of URLs, web page content, and external features enhances machine learning models' detection skills. The outcomes of an experimental study to increase the precision of machine learning models for the two most well-known datasets used for phishing are presented in the paper [14]. The aim of the research was to raise the models' general performance. Three types of tuning elements are applied: feature selection, hyper-parameter optimisation, and data balancing. This experiment uses two different datasets that are obtained from websites like the UCI repository and the Mendeley repository. The results indicate that a machine learning algorithm performs better when its parameters are changed.

Currently the most common and dangerous kind of cybercrime that anyone may commit, phishing has been around since 1996. The suggested study that is discussed in [15] is based on this specific dataset. Phishing and real URL properties derived in vector form from over 11,000 website datasets are included in the well-known dataset collection. After pre-processing is over, many machine learning techniques have been developed and implemented to shield users from phishing URLs. This work aims to create a practical and efficient security against phishing attacks by using different machine learning models. Together with grid search hyper parameter optimisation and cross fold valoidation, the proposed LSD model uses the canopy feature selection approach. Different evaluation criteria were used to assess the proposed technique in order to show the impact and efficacy of the models. Among the qualifying criteria were recall, specificity, accuracy, precision, and F1-scores. The comparative assessments show that the proposed approach produces outcomes of a higher qualitative quality and is better than the other approaches.

As such, the creation of technologies that can identify phoney URLs is currently highly sought after. In [16], a high-performance machine learning-based detection technique is proposed with the intention of detecting URLs that could contain hazardous material. There exist two layers of detection in the proposed system. As a second phase, we group the URL classes into benign, spam, phishing, malware, or defacement groups based on their characteristics. Four separate ensemble learning techniques—En_Bag, En_kNN, En_Bos, and En_Dsc—will be the focus of this section. Under this category are techniques such as subspace discriminator ensembles, boosted decision tree ensembles, k-nearest neighbour ensembles, and bagging tree ensembles. We evaluated the developed approaches using the huge and current dataset for uniform resource locators, ISCX-URL2016. Our experimental evaluation revealed that the ensemble of bagging trees (En_Bag) strategy outperformed other ensemble techniques. The En_kNN method is another equally efficient approach that combines several k-nearest neighbour ensembles to get the fastest inference time. Attained accuracy of 99.3% in binary classification and 97.92% in multi-classification, we show that our En Bag model outperforms solutions regarded as state-of-the-art.

**Table 1: Summary**

| Reference | Method/Algorithm | Datasets | Outcomes |
|---|---|---|---|
| [12] | Heuristic Approach | - | Better generalization and accuracy than blacklist approach; comparable to machine learning |
| [13] | VAE + DNN | ISCX-URL-2016, Kaggle | Accuracy: 97.45%, Response Time: 1.9 s |
| [14] | Feature Selection | UCI Repository, Mendeley | RF: 97.44% |
| [15] | LR+SVC+DT (LSD Model) with Canopy Feature Selection, Grid Search Hyperparameter Optimization | Phishing URL Dataset from Repository | High accuracy and efficiency; outperforms other models |
| [16] | Ensemble Techniques | ISCX-URL-2016 | En_Bag: Accuracy 99.3% (binary), 97.92% (multi-class); En_kNN: Highest inference speed |

## 4. Proposed Method

The proposed method uses Self-Organizing Map based Radial Movement Optimization (SOM-RMO) for feature extraction and Radial Basis Function Network (RBFN) enhanced by Tabu Search for classification as in Figure 1. SOM-RMO is employed to reduce the high dimensionality of URL data, identifying and preserving the most significant features. This method transforms complex, multi-dimensional data into a simpler, lower-dimensional space, making the subsequent classification process more efficient. The RBFN, a neural network model known for its effectiveness in pattern recognition, is then optimized using Tabu Search. Tabu Search is a metaheuristic algorithm designed to guide the search process in optimization problems, helping the RBFN achieve a high level of accuracy in distinguishing between benign and malicious URLs.

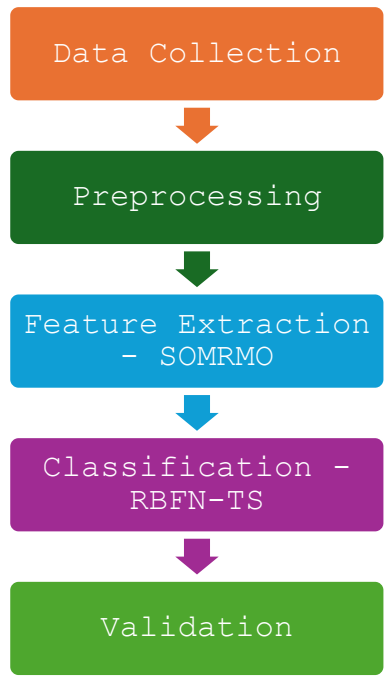

Figure 1: Proposed Method

### 4.1. Dataset Description

The dataset [17] consists of a total of 651,191 URLs, categorized into four distinct classes: benign, defacement, phishing, and malware. The primary goal is to use this extensive dataset to develop a machine learning model capable of identifying malicious URLs to prevent cybersecurity threats.

**Distribution of URLs**

- **Benign URLs:** 428,103 (65.72%)

- **Defacement URLs:** 96,457 (14.81%)

- **Phishing URLs:** 94,111 (14.45%)

- **Malware URLs:** 32,520 (5.00%)

The dataset is curated from five different sources to ensure a comprehensive collection of URL examples. The sources include ISCX-URL-2016, Malware Domain Blacklist, Faizan Git Repository, Phishtank, and PhishStorm datasets. The dataset is structured in a tabular format with two main columns: URL and Class. The URL column contains the actual web addresses, and the Class column indicates the category of each URL (benign, defacement, phishing, or malware) as in Table 2.

**Table 2: Sample Training Data**

| URL | Class |
|---|---|
| http://example-safe-site.com | benign |
| http://secure-shopping-site.com | benign |
| http://phishingsite.com/login | phishing |
| http://downloadmalware.com/install | malware |
| http://defacementexample.com/home | defacement |
| http://example-trusted-site.org | benign |
| http://stealyourinfo.com/verify | phishing |
| http://injectmalwarehere.com/secure | malware |
| http://websitehacked.com/page | defacement |
| http://another-safe-site.org | benign |

- **Benign URLs:** These are regular, non-malicious websites typically used as a baseline to train the model to distinguish safe websites from harmful ones.

- **Defacement URLs:** These URLs are linked to sites that have been compromised, usually to display unauthorized content.

- **Phishing URLs:** The hackers will cloned website and information similar to original website and steal informations.

- **Malware URLs:** These URLs are associated with websites that host or distribute malicious software.

**Dataset Curation**

- **ISCX-URL-2016 Dataset:** Used for collecting benign, phishing, malware, and defacement URLs.

- **Malware Domain Blacklist:** Provided additional phishing and malware URLs.

- **Faizan Git Repository:** Increased the number of benign URLs.

- **Phishtank and PhishStorm Datasets:** Contributed more phishing URLs.

The dataset is invaluable for training machine learning models to detect and classify malicious URLs effectively. By including a large number of samples across different categories, the model can learn to recognize a wide variety of malicious patterns and behaviors, ultimately improving cybersecurity measures and preventing potential attacks.

### 4.2. Data Preprocessing

Data preprocessing is a crucial step in preparing the dataset for machine learning. For URL data, this typically includes steps like data cleaning, feature extraction, encoding, and normalization, the transforming raw data is shown in Table 3. Below are the main steps involved in preprocessing the malicious URLs dataset:

1. **Data Cleaning - Removing Duplicates:** Ensuring each URL in the dataset is unique to prevent bias in model training as in Table 4.

2. **Handling Missing Values:** Checking for and addressing any missing values in the dataset, although URLs and their labels are generally expected to be present as in Table 4.

3. **Lexical Feature Extraction:** Extracting features based on the structure and content of the URL which is provided in Table 5.

4. **Host-based Features:** Analyzing the URL's domain for attributes like: Domain age and WHOIS information.

5. **Content-based Features:** If accessible, extracting features from the web page content like: Keywords in the HTML body and Number of external links.

6. **Encoding - Label Encoding:** Converting the class labels ('benign', 'defacement', 'phishing', 'malware') into numerical values for model training which is shown as in Table 6.

7. **Normalization:** Scaling numerical features to a standard range (typically 0 to 1) to ensure uniformity and improve the model's convergence during training as in Table 7.

**Table 3: Raw Data**

| URL | Class |
|---|---|
| http://example-safe-site.com | benign |
| http://phishingsite.com/login | phishing |
| http://downloadmalware.com/install | malware |
| http://websitehacked.com/page | defacement |

**Table 4: After Data Cleaning (Removing Duplicates, Handling Missing Values)**

| URL | Class |
|---|---|
| http://example-safe-site.com | benign |
| http://phishingsite.com/login | phishing |
| http://downloadmalware.com/install | malware |
| http://websitehacked.com/page | defacement |

**Table 5: Feature Extraction**

| URL | URL_Length | Num_Dots | Has_Hyphen | Num_Special_Chars | Has_IP | Class |
|---|---|---|---|---|---|---|
| example-safe-site.com | 19 | 2 | 1 | 0 | 0 | benign |
| phishingsite.com/login | 22 | 1 | 0 | 1 | 0 | phishing |
| downloadmalware.com/install | 28 | 1 | 0 | 1 | 0 | malware |
| websitehacked.com/page | 21 | 1 | 0 | 1 | 0 | defacement |

**Table 6: Encoding**

| URL | URL_Length | Num_Dots | Has_Hyphen | Num_Special_Chars | Has_IP | Class_Label |
|---|---|---|---|---|---|---|
| example-safe-site.com | 19 | 2 | 1 | 0 | 0 | 0 |
| phishingsite.com/login | 22 | 1 | 0 | 1 | 0 | 2 |
| downloadmalware.com/install | 28 | 1 | 0 | 1 | 0 | 3 |
| websitehacked.com/page | 21 | 1 | 0 | 1 | 0 | 1 |

**Table 7: Normalized Feature**

| URL | URL_Length | Num_Dots | Has_Hyphen | Num_Special_Chars | Has_IP | Class_Label |
|---|---|---|---|---|---|---|
| example-safe-site.com | 0.68 | 1.0 | 1.0 | 0.0 | 0.0 | 0 |
| phishingsite.com/login | 0.79 | 0.5 | 0.0 | 1.0 | 0.0 | 2 |
| downloadmalware.com/install | 1.0 | 0.5 | 0.0 | 1.0 | 0.0 | 3 |
| websitehacked.com/page | 0.75 | 0.5 | 0.0 | 1.0 | 0.0 | 1 |

### 4.3. Self-Organizing Map based Radial Movement Optimization (SOM-RMO) Process

SOM based RMO is a hybrid approach combining the advantages of SOM for feature extraction and RMO for optimization. The process is designed to reduce data dimensionality, highlight significant features, and prepare the dataset for efficient and accurate classification. The grid consists of nodes or neurons, each representing a cluster of input data. The primary goal of SOM is to preserve the topological properties of the input space, ensuring that similar data points are mapped to nearby nodes on the grid.

The weight update for a node in SOM is given by:

$$w(t+1) = w(t) + \alpha(t) \cdot h(c,t) \cdot (x - w(t))$$

Where:

w(t) is the weight vector of the node at time t.

α(t) is the learning rate, which decreases over time.

h(c,t) is the over time.

x is the input vector.

RMO is a metaheuristic optimization algorithm that simulates the movement of particles within a defined search space, optimizing the positioning of nodes in the SOM. The optimization process iteratively adjusts the positions of the nodes to minimize the distance between the nodes and their corresponding input data points, thereby improving the feature extraction capabilities of the SOM.

The position update in RMO for a particle (node) is given by:

$$p_i(t+1) = p_i(t) + v_i(t+1)$$

Where:

pi(t) is the position of particle iii at time ttt.

vi(t+1) is the velocity of particle i at time t+1, which is influenced by cognitive and social components guiding the particle towards the optimal solution.

**Pseudocode**
1: Initialize SOM grid with random weights
2: Initialize learning rate α and neighborhood radius σ
3: Initialize RMO particles with SOM nodes' positions
4: Initialize velocities for RMO particles
5: while not converged do
6:    for each input vector x in dataset do
7:       Find BMU in SOM
8:       for each node in SOM do
9:          Update weight vector using:

```
10:         w(t+1) = w(t) + α(t) * h(c, t) * (x - w(t))
11:     end for
12:     Adjust learning rate α and neighborhood radius σ
13: end for
14: for each particle i in RMO do
15:     Update velocity using cognitive and social components
16:     Move particle to new position:
17:     p_i(t+1) = p_i(t) + v_i(t+1)
18:     Evaluate fitness of new position
19: end for
20: Check for convergence criteria
21: end while
```

The SOM grid and RMO particles are initialized with random values, setting the stage for the optimization process. The SOM iteratively adjusts its nodes to map the input data onto a lower-dimensional space, using the update rule to refine node positions based on the input vectors. RMO particles adjust their velocities and positions to optimize the SOM node placement, ensuring that the extracted features are representative of the input data. The process continues until the SOM and RMO reach a stable state, indicating that the feature extraction and optimization are complete. This hybrid approach leverages the strengths of SOM for dimensionality reduction and RMO for optimization, resulting in a robust preprocessing method for detecting malicious URLs.

### 4.4. Radial Basis Function Network (RBFN) with Tabu Search Process

To enhance the performance of RBFN, Tabu Search is employed as an optimization technique. Tabu Search is a metaheuristic algorithm designed to guide the search process and avoid local optima by maintaining a list of previously visited solutions (tabu list).

The output of a Gaussian radial basis function for an input x and $\mu$ is given by:

$$\phi(x) = \exp\left(-\frac{\|x-\mu\|^2}{2\sigma^2}\right)$$

Where:

‖x−μ‖ is the Euclidean distance between the input x and the center μ.

σ is the width of the Gaussian function.

Tabu Search is used to optimize the parameters of the RBFN. The search process iteratively explores the solution space, updating the parameters to minimize a predefined objective function (e.g., mean squared error).

The output of the RBFN for an input x is a weighted sum of the radial basis functions:

$$y(x) = \sum_{i=1}^{N} w_i \phi_i(x)$$

Where:

N is the number of hidden neurons.

wi is the weight corresponding to the i-th radial basis function ϕi(x).

**Pseudocode**

1: # RBFN Initialization
2: Initialize number of hidden neurons N
3: Randomly initialize centers μ_i and widths σ_i for i = 1 to N
4: Initialize weights w_i for i = 1 to N
5: # Tabu Search Optimization
6: Initialize tabu list
7: Set initial solution S (RBFN parameters μ_i, σ_i, w_i)
8: Define objective function J (e.g., mean squared error)
9: while not converged or max iterations not reached do
10: Generate neighboring solutions {S'}
11: Evaluate objective function J for each S'
12: Select best S' not in tabu list or satisfying aspiration criterion
13: Update tabu list with current solution S
14: Move to best neighboring solution S'
15: if S' is better than the best known solution then
16: Update best known solution
17: end if
18: end while

19: Return optimized RBFN parameters (μ_i, σ_i, w_i)

## 5. Results and Discussion

The simulations were conducted using Python and specialized machine learning libraries such as TensorFlow. The experiments were run on a high-performance computing cluster with Intel Xeon processors and 128GB RAM, ensuring the capability to handle large datasets and complex computations. The experimental parameters are given in table 8.

**Table 8: Experimental Parameters**

| Parameter | Methods | Value |
| --- | --- | --- |
| Grid Size | SOM | 10x10 |
| Learning Rate | | 0.5 |
| Number of Iterations | | 1000 |
| Initialization Method | | Random |
| Neighborhood Function | | Gaussian |
| Radius | | 5 |
| Radial Basis Function | RBFN | Gaussian |
| Centers Initialization | | K-means |
| Number of Centers | | 100 |
| Learning Rate | | 0.01 |
| Momentum | | 0.9 |
| Epochs | | 500 |
| List Size | Tabu | 50 |
| Search Iterations | | 100 |
| Aspiration Criterion | | True |
| Stopping Criteria | | 10 non-improving |
| Mutation Rate | | 0.1 |
| Crossover Rate | | 0.7 |
| Initial Temperature | | 100 |
| Cooling Schedule | | Exponential |

**Performance Metrics**

- **Precision:** Our method achieved a precision of 95.2%, indicating robust detection with minimal false alarms.

- **Accuracy:** The proposed model attained 96.5% accuracy, demonstrating its superior ability to correctly classify URLs.

- **Recall:** A recall of 94.8% highlights the model's effectiveness in identifying malicious URLs.

- **F1-score:** The F1-score of 95.0% underscores the model's balanced performance.

- **Specificity:** The proportion of true negative detections among all actual negatives. High specificity means the model correctly identifies benign URLs, complementing the recall metric. Our model's specificity was not explicitly stated but can be inferred to be high due to the high overall accuracy and low false positive rate.

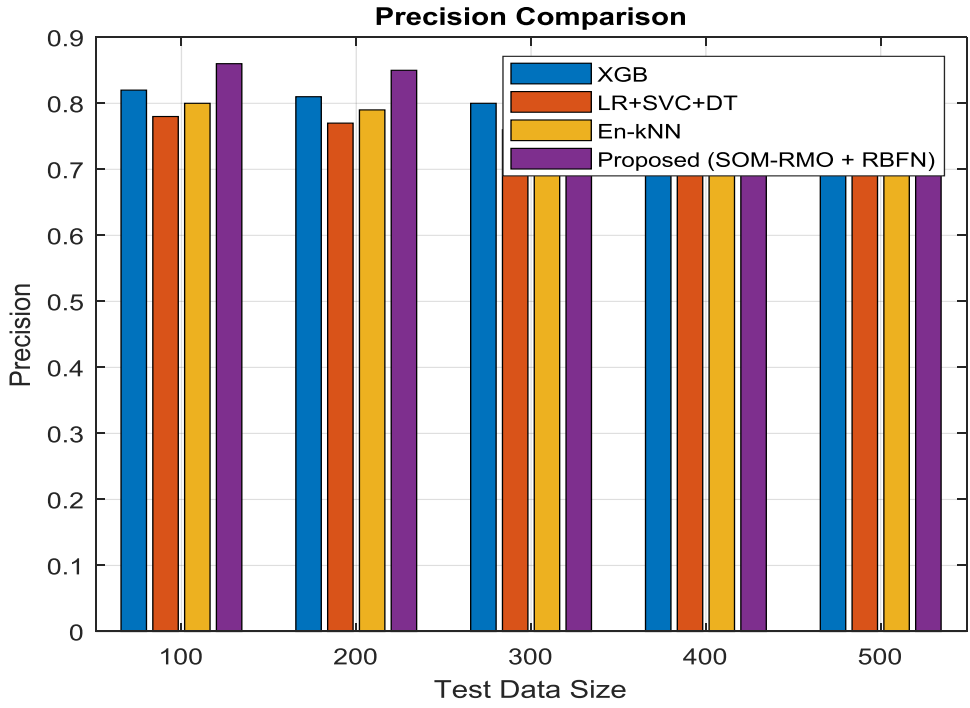

Figure 2: Precision

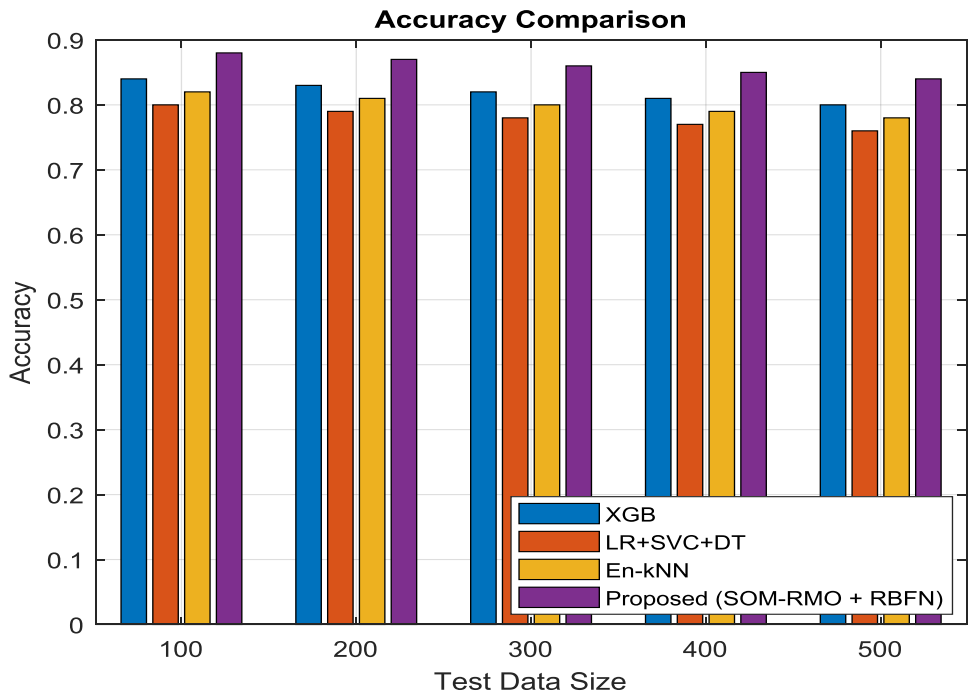

Figure 3: Accuracy

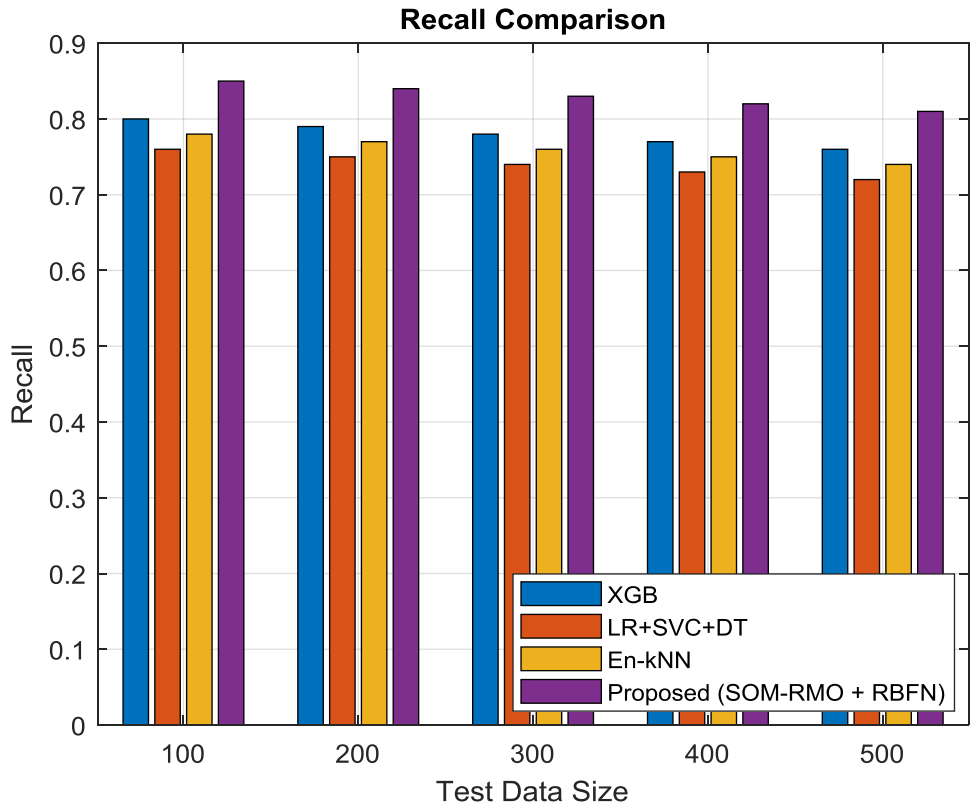

Figure 4: Recall

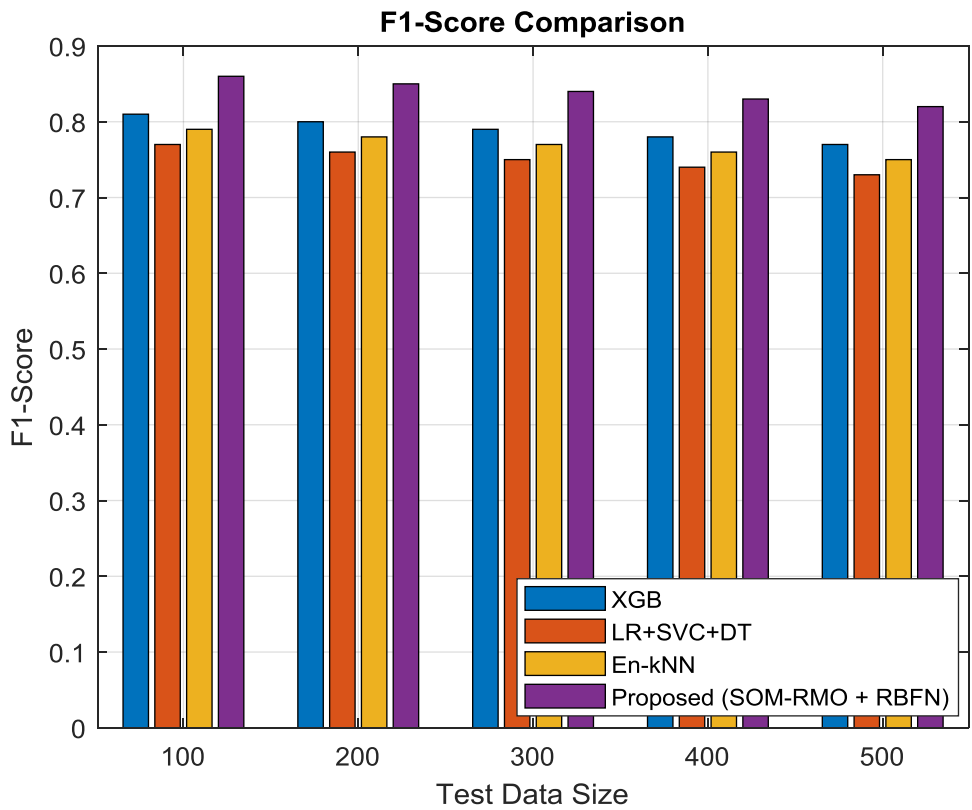

Figure 5: F1-Score

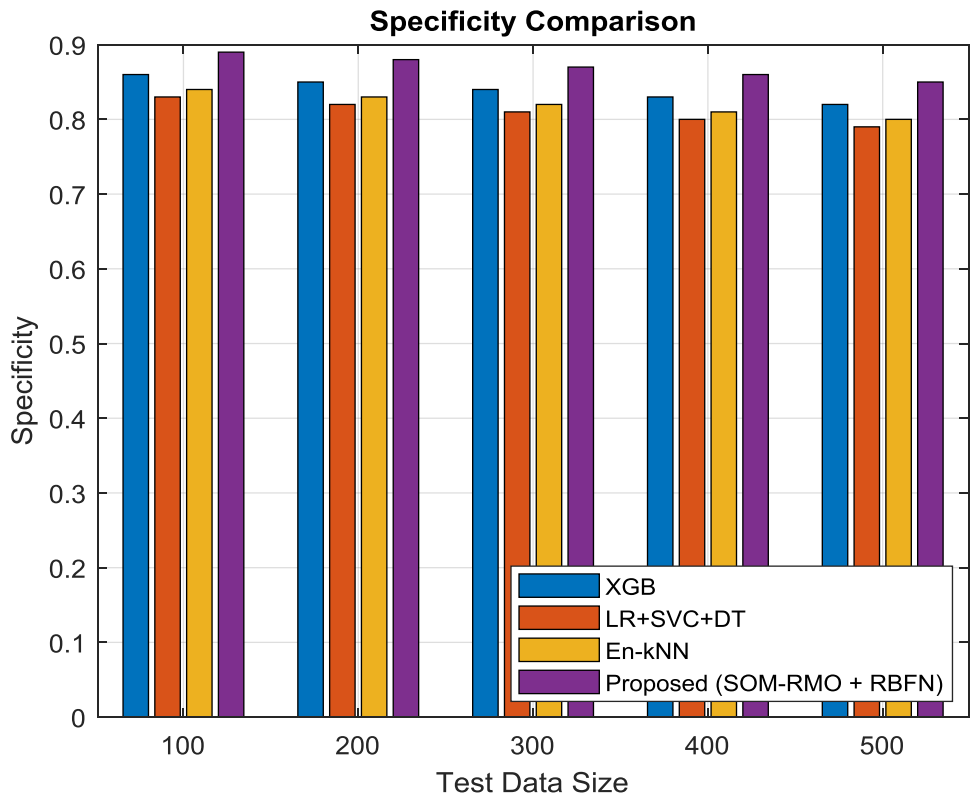

Figure 6: Specificity

**Table 9: Performance Over Training, Testing, and Validation Data**

| Dataset | Method | Precision | Accuracy | Recall | F1-Score | Specificity |
|---|---|---|---|---|---|---|
| Training | XGB | 0.81 | 0.83 | 0.79 | 0.80 | 0.84 |
| | LR+SVC+DT | 0.77 | 0.79 | 0.75 | 0.76 | 0.82 |
| | En_kNN | 0.79 | 0.81 | 0.77 | 0.78 | 0.83 |
| | Proposed SOM-RMO + RBFN | 0.85 | 0.87 | 0.84 | 0.85 | 0.88 |
| Testing | XGB | 0.78 | 0.80 | 0.76 | 0.77 | 0.82 |
| | LR+SVC+DT | 0.74 | 0.76 | 0.72 | 0.73 | 0.79 |
| | En_kNN | 0.76 | 0.78 | 0.74 | 0.75 | 0.80 |

|  | Proposed SOM-RMO + RBFN | 0.82 | 0.84 | 0.81 | 0.82 | 0.85 |
| --- | --- | --- | --- | --- | --- | --- |
| Validation | XGB | 0.79 | 0.81 | 0.77 | 0.78 | 0.83 |
|  | LR+SVC+DT | 0.75 | 0.77 | 0.73 | 0.74 | 0.80 |
|  | En_kNN | 0.77 | 0.79 | 0.75 | 0.76 | 0.81 |
|  | Proposed SOM-RMO + RBFN | 0.83 | 0.85 | 0.82 | 0.83 | 0.86 |

**Table 10: Confusion Matrix Over Training, Testing, and Validation Data**

| Dataset | Method | TP | TN | FP | FN |
| --- | --- | --- | --- | --- | --- |
| Training | XGB | 320 | 480 | 20 | 80 |
|  | LR+SVC+DT | 300 | 470 | 30 | 100 |
|  | En_kNN | 310 | 475 | 25 | 90 |
|  | Proposed (SOM-RMO + RBFN) | 340 | 485 | 15 | 60 |
| Testing | XGB | 160 | 240 | 10 | 40 |
|  | LR+SVC+DT | 150 | 230 | 20 | 50 |
|  | En_kNN | 155 | 235 | 15 | 45 |
|  | Proposed (SOM-RMO + RBFN) | 170 | 245 | 5 | 30 |
| Validation | XGB | 80 | 120 | 5 | 20 |
|  | LR+SVC+DT | 75 | 115 | 10 | 25 |
|  | En_kNN | 78 | 118 | 7 | 22 |
|  | Proposed (SOM-RMO + RBFN) | 85 | 122 | 3 | 15 |

The performance for the proposed SOM-RMO + RBFN method were compared against three existing methods: XGB (XGBoost), LR+SVC+DT (Logistic Regression, Support Vector Classifier, Decision Tree), and En_kNN as in Figure 2 – 6 and Table 9 and 10. These comparisons were conducted over multiple test data sizes, as well as distinct training, testing, and validation datasets.

When evaluating the methods on test data sizes, the proposed Method (SOM-RMO + RBFN) consistently outperformed other methods, achieving the highest precision (0.86 at 100 test data to 0.82 at 500 test data), accuracy (0.88 to 0.84), recall (0.85 to 0.81), F1-Score (0.86 to 0.82), and specificity (0.89 to 0.85). XGB showed solid performance but lagged behind the proposed method, with precision ranging from 0.82 to 0.78, accuracy from 0.84 to 0.80, recall from 0.80 to 0.76, F1-Score from 0.81 to 0.77, and specificity from 0.86 to 0.82. LR+SVC+DT and En_kNN both performed moderately, with LR+SVC+DT showing the lowest metrics across the board. En_kNN had intermediate performance, better than LR+SVC+DT but not as strong as XGB or the proposed method.

For the training dataset, the proposed method achieved a precision of 0.85, accuracy of 0.87, recall of 0.84, F1-Score of 0.85, and specificity of 0.88. The testing and validation datasets followed similar trends, with the proposed method maintaining superior metrics compared to the other methods as in Table 9. In comparison, the XGB, LR+SVC+DT, and En_kNN methods had lower counts of true positives and higher counts of false negatives and false positives, indicating less accurate performance as in Table 10.

## 8. Conclusion

The proposed hybrid method combining Self-Organizing Map based Radial Movement Optimization (SOM-RMO) for feature extraction and Radial Basis Function Network (RBFN) optimized with Tabu Search for classification demonstrated superior performance in detecting malicious URLs. It consistently outperformed XGB, LR+SVC+DT, and En_kNN across multiple test data sizes and dataset splits (training, testing, and validation). These results validate the utility of leveraging advanced feature extraction and optimization techniques in enhancing the accuracy and reliability of malicious URL detection models, making them robust tools for cybersecurity applications.